\documentclass[prl,aps,twocolumn,showpacs,floatfix,amsmath,amssymb,aps]{revtex4-2}
\usepackage{amsfonts}
\usepackage{amssymb}
\usepackage[linesnumbered,ruled,vlined]{algorithm2e}
\usepackage{hyperref}
\usepackage{graphicx}
\usepackage{dcolumn}
\usepackage{bm,amsmath,verbatim}
\usepackage{mathrsfs}
\usepackage{color}
\usepackage{mathtools}
\usepackage{amsmath,amssymb,amsthm}

\usepackage[T1]{fontenc}
\usepackage[latin9]{inputenc}
\usepackage{booktabs}
\providecommand{\tabularnewline}{\\}

\hypersetup{colorlinks,linkcolor=blue,citecolor=blue,filecolor=blue,urlcolor=blue}

\begin{document}
\title{Self-dual Stacked Quantum Low-Density Parity-Check Codes}
\author{Ze-Chuan Liu$^{1}$}
\thanks{These authors contribute equally to this work.}
\author{Chong-Yuan Xu$^{1}$}
\thanks{These authors contribute equally to this work.}
\author{Yong Xu$^{1,2}$}
\email{yongxuphy@tsinghua.edu.cn}
\affiliation{$^{1}$Center for Quantum Information, IIIS, Tsinghua University, Beijing 100084, People's Republic of China}
\affiliation{$^{2}$Hefei National Laboratory, Hefei 230088, People's Republic of China}

\begin{abstract}
Quantum low-density parity-check (qLDPC) codes are promising candidates for 
fault-tolerant quantum computation due to their high encoding rates and distances.
However, implementing logical operations using qLDPC codes presents significant 
challenges. Previous research has demonstrated that self-dual qLDPC codes 
facilitate the implementation of transversal Clifford gates.
Here we introduce a method for constructing self-dual qLDPC codes by 
stacking non-self-dual qLDPC codes. Leveraging this methodology, we develop 
double-chain bicycle codes, double-layer bivariate bicycle (BB) codes, 
double-layer twisted BB codes, and double-layer reflection codes, many of 
which exhibit favorable code parameters. Additionally, we conduct numerical 
calculations to assess the performance of these codes as quantum memory under the 
circuit-level noise model, revealing that the logical failure rate can be 
significantly reduced with high pseudo-thresholds.
\end{abstract}

\maketitle
Given the inevitability of noise in quantum information processing, 
it is essential to utilize methods to mitigate its effects. Quantum error-correcting 
codes offer a feasible and reliable approach for achieving fault-tolerant 
quantum computation~\cite{shor1995scheme,steane1996multiple,calderbank1996good,lafiamme1996perfect,kitaev2003fault,panteleev2022asymptotically,leverrier2022quantum,xu2024constant,bravyi2024high}. A prominent example is the surface code, 
which features local stabilizers and straightforward decoding procedures 
and has been experimentally demonstrated on quantum devices~\cite{zhao2022realization,google2023suppressing,bluvstein2024logical,google2025quantum,eickbusch2025demonstration}. 
However, due to the low encoding rate of the surface code, implementing 
large-scale quantum computation incurs substantial resource overhead. 
Recently, qLDPC codes~\cite{tillich2013quantum,Gottesman2013,kovalev2013quantum,breuckmann2016constructions,panteleev2021degenerate,breuckmann2021balanced,krishna2021fault,higgott2021subsystem,breuckmann2021quantum,panteleev2021quantum,delfosse2021bounds,baspin2022connectivity,baspin2022quantifying,tremblay2022constant,cohen2022low,panteleev2022asymptotically,leverrier2022quantum,strikis2023quantum,quintavalle2023partitioning,lin2024quantum,xu2024constant,bravyi2024high,zhang2025time,li2025low}, such as the BB code~\cite{bravyi2024high}, have been 
introduced for fault-tolerant quantum computation. These codes provide a 
high encoding rate while maintaining a large code 
distance~\cite{bravyi2024high}, making them promising candidates for low-overhead 
fault-tolerant quantum computation. In recent years, there has been 
growing interest in investigating their structure, performance, and 
potential extensions~\cite{liang2024extracting,liang2024operator, liang2025generalized, liang2025self,wang2026demonstration}.

Despite the advantages of qLDPC codes, 
implementing logical gates remains a significant challenge. Existing methods, 
such as homomorphic measurements~\cite{huang2023homomorphic,xu2025fast,ide2025fault} and generalized lattice surgery~\cite{cohen2022low,cross2024improved,cowtan2025parallel}, 
often require additional ancilla qubits and complex error correction procedures. 
Therefore, it is desirable to identify qLDPC codes that allow for transversal 
logical gate constructions. Recent studies indicate that self-dual qLDPC codes 
permit transversal Clifford gates~\cite{tansuwannont2025clifford,liang2025self,reddy2026asymptotically}. Moreover, self-dual qLDPC codes 
can be utilized to construct fermionic codes for fermionic quantum computation based 
on Majorana modes~\cite{schuckert2024fermion,ott2024error,Mudassar2025arxiv,xu2025fermion}. While some self-dual qLDPC codes, such as bicycle codes~\cite{MacKay2004ieee} and 
certain BB codes~\cite{bravyi2024high,liang2025self}, exist, a general framework for constructing a self-dual code 
from a non-self-dual code remains undeveloped.

Here, we construct a class of self-dual Calderbank-Shor-Steane (CSS) LDPC codes
by stacking non-self-dual CSS codes.
We examine examples of double-chain bicycle codes, double-layer BB codes, 
double-layer twisted BB codes, and double-layer reflection codes. 
We calculate the parameters for each type of code and find that many exhibit odd-weight
logical operators and exhibit favorable code parameters.  
Finally, we perform numerical simulations for several codes as qubit memory 
and observe that the logical failure rate can be significantly reduced, 
with the pseudo-threshold exceeding $0.7\%$ under the circuit-level noise
model.

\begin{figure}
\includegraphics[width=1\linewidth]{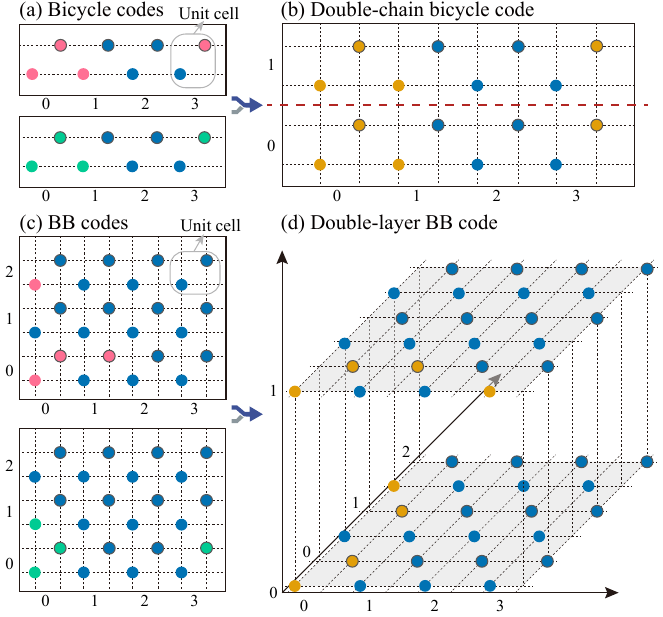}
\caption{Illustration of a double-chain bicycle code and a double-layer BB code
obtained from a bicycle code with $A=I_4+S_4^3$ and $B=I_4+S_4$ 
and a BB code with $A=I_{12}+T_x$ and $B=I_{12}+T_y^2$, respectively. 
In (a) and (c), the support of an $X$ ($Z$) stabilizer generator
is highlighted by filled pink (green) circles in
the upper (lower) panel.
In (b) and (d), the support of an $X$ (or $Z$) stabilizer generator
is indicated
by filled orange circles.
}
\label{fig1}
\end{figure}

\emph{Code construction}.---We start by introducing our method to construct
a self-dual CSS code based on a base CSS code that is not required to be
self-dual. This base code is described by the $X$ and $Z$ stabilizer matrices,
$h_X$ and $h_Z$, expressed as 
$h_X=\begin{pmatrix}
A & B
\end{pmatrix}$ and 
$h_Z=\begin{pmatrix}
B^T & A^T
\end{pmatrix}$,
where $A$ and $B$ are square binary matrices.
We require that $[A,B]=0$, $[A,A^T]=0$, and $[B,B^T]=0$ so that 
$h_X h_Z^T=0$, forming a well-defined code.
We use the matrices $A$ and $B$ to construct new check matrices, 
\begin{equation}
H_X=H_Z = 
\begin{pmatrix}
A & B^T & A^T & B\\
B^T & A & B & A^T\\
\end{pmatrix}\\
= 
\begin{pmatrix}
U & U^T 
\end{pmatrix},
\end{equation}
where 
$
U=I_2\otimes A+\sigma_x\otimes B^T
$
with $I_l$ being an $l\times l$ identity matrix and $\sigma_x$ being
a Pauli matrix.
It follows that $H_X H_Z^T = 0$ since $U U^T=U^T U$, thus obtaining a self-dual CSS code.
If the base code is a qLDPC code, then the stacked code remains a qLDPC code, as
the weight of stacked code's stabilizer generators is at most double 
maximum weight of the base code's stabilizer generators.
We will use the bicycle code, BB code, and twisted BB code as base codes 
to construct self-dual codes. Finally, we propose a new class of base codes 
by introducing reflection operations to construct self-dual codes.

\emph{Double-chain bicycle code}.---We first use the bicycle code~\cite{MacKay2004ieee} on 
a 1D periodic chain of length $l$ to construct a double-chain bicycle code.
In this setting, the matrices $A$ and $B$ are chosen to be circulant~\cite{MacKay2004ieee}.
Specifically, $A$ and $B$ are expressed as
$A = \sum_{j=0}^{l-1} a_j S_l^j$ 
and $B = \sum_{j=0}^{l-1} b_j S_l^j$,
where $a_j, b_j \in \mathbb{Z}_2$ and $S_l$ is the $l \times l$ cyclic shift matrix,
whose only nonzero entry in row $j$ appears in column $(j+1) \bmod l$.
Clearly, $A$ and $B$ satisfy that $[A,B]=[A,A^T]=[B,B^T]=0$, 
allowing the code to be used as a base code to construct 
self-dual double-chain bicycle codes.
A base code's physical layout is obtained by placing each qubit
at each lattice site for a lattice obtained by translating a 
unit cell consisting of two sites, as illustrated in Fig.~\ref{fig1}(a).
Given $A$ and $B$, an $X$ ($Z$) stabilizer generator for the base code 
called seed stabilizer is given by 
$s_X=\prod_{j} X_{0,j}^{a_j}\prod_{j'} X_{1,j'}^{b_{j'}}$
($s_Z=\prod_{j} Z_{0,l-j}^{b_j}\prod_{j'} Z_{1,l-j'}^{a_{j'}}$),
where $X_{\nu,j}$ ($Z_{\nu,j}$) denotes the Pauli $X$ ($Z$) operator 
at the $\nu$th site of the $j$th unit cell.
All other generators are produced by translating this
stabilizer while considering periodic boundary conditions (PBCs) 
[see Fig.~\ref{fig1}(a) for an example].
For the self-dual code, two chains are included with the first and third block columns 
in the check matrix corresponding to the first chain
and the second and fourth block columns corresponding to the second chain.
The former and latter two block columns correspond to the
upper and lower site in a unit cell, respectively.
An $X$ stabilizer generator is expressed as 
$S_X=\prod_{j_1} X_{0,(j_1,0)}^{a_{j_1}}\prod_{j_2} X_{0,(l-j_2,1)}^{b_{j_2}}
\prod_{j_3} X_{1,(l-j_3,0)}^{a_{j_3}}\prod_{j_4} X_{1,(j_4,1)}^{b_{j_4}}$,
where in $(j,j_z)$, $j$ and $j_z$ are the unit cell index and  
the chain index, respectively. An example is provided in
Fig.~\ref{fig1}(b).

Table~\ref{table1} lists ten representative weight-eight 
double-chain bicycle codes (see more in Table~\ref{tableEM1} in End Matter) with parameters $[[n,k,d]]$ 
by considering  
$A=S_l^\alpha+S_l^\beta$ and $B=S_l^\gamma+S_l^\delta$,
where $n$, $k$, and $d$ denote the number of physical 
qubits, the number of logical qubits, and code distance,
respectively. 
These codes are obtained by numerically searching
over various lattice sizes $l$ and  
different parameters $\alpha$, $\beta$, $\gamma$, and $\delta$.
The results indicate a high encoding rate. In addition,
we observe that when $l$ is even, it is highly probable that
the code is an even code, meaning that the weight of logical operators
cannot be odd. Conversely, when $l$ is odd, the opposite holds true.

\begin{table}
\begin{ruledtabular}
\centering
\caption{Weight-eight double-chain bicycle codes $[[n,k,d]]$  
constructed from weight-four bicycle codes with 
$A=x^\alpha+x^\beta$ and $B=x^\gamma+x^\delta$, where $x=S_l$.
Code distances are computed using the integer programming approach~\cite{bravyi2024high,landahl2011fault}.
The top five codes have logical operators of odd-weight, while
the bottom five codes can only have logical operators of even-weight 
(the same holds for Table~\ref{table2}--Table~\ref{table4}).
Here, $1$ represents an identity matrix.}
\label{table1}
\begin{tabular}{c c c c}
$[[n,k,d]]$ & $l$ & $A$ & $B$  \tabularnewline
\midrule		
$[[36,4,6]]$ & 9 & $1+x^4$ & $x^3+x^6$  \tabularnewline		
$[[84,12,6]]$ & 21 & $x^2+x^5$ & $x^5+x^{14}$  \tabularnewline	
$[[100,12,8]]$ & 25 & $x^{10}+x^{24}$ & $x^{10}+x^{16}$  \tabularnewline
$[[108,4,12]]$ & 27 & $x^{22}+x^{24}$ & $x^{12}+x^{22}$  \tabularnewline
$[[132,8,12]]$ & 33 & $x^{10}+x^{11}$ & $x^{11}+x^{31}$  \tabularnewline
\midrule
$[[24,8,4]]$ & 6 & $1+x^2$ & $x^3+x^4$  \tabularnewline
$[[72,6,8]]$ & 18 & $x^2+x^{17}$ & $x^4+x^5$ \tabularnewline
$[[80,8,8]]$ & 20 & $1+x^{17}$ & $x^8+x^{17}$  \tabularnewline
$[[88,4,10]]$ & 22 & $x^{13}+x^{18}$ & $x+x^5$  \tabularnewline
$[[104,6,12]]$ & 26 & $x^6+x^{11}$ & $x^5+x^{14}$ \tabularnewline
\end{tabular}
\end{ruledtabular}
\end{table}

\emph{Double-layer BB code}.---The BB code is a generalization 
of the bicycle code by considering 
$A = \sum_{j_x,j_y} a_{j_x j_y} T_x^{j_x} T_y^{j_y}$
and $B = \sum_{j_x,j_y} b_{j_x j_y} T_x^{j_x} T_y^{j_y}$,
where 
$T_x = S_l \otimes I_m$ and
$T_y = I_l \otimes S_m$ with $l$ and $m$ being positive integers,
and $a_{j_x j_y},b_{j_x j_y}\in \mathbb{Z}_2$ with 
$j_x$ and $j_y$ running from
$0$ to $l-1$ and $m-1$, respectively~\cite{bravyi2024high}.
When $m=1$ or $l=1$, the code reduces to the bicycle code.
The matrices $A$ and $B$ satisfy the required 
commutation relation for constructing the self-dual double-layer 
BB code.
Anagolous to the bicycle code,
the physical arrangement of the BB codes corresponds to translating
a unit cell comprising two qubits in the $x$ and $y$
directions, as illustrated in Fig.~\ref{fig1}(c).
For the matrices $A$ and $B$ presented above, the $X$ ($Z$) 
seed stabilizer of the BB code is expressed as
$s_X=\prod_{j_x, j_y} X_{0,(j_x,j_y)}^{a_{j_x j_y}}\prod_{j_x', j_y'} X_{1,(j_x',j_y')}^{b_{j_x' j_y'}}$
($s_Z=\prod_{j_x, j_y} Z_{0,(j_x,j_y)}^{a_{j_x j_y}}\prod_{j_x', j_y'} Z_{1,(j_x',j_y')}^{b_{j_x' j_y'}}$),
where the subscript $\nu,(j_x,j_y)$ denotes the qubit at the $\nu$th site in a unit cell labeled by
$(j_x,j_y)$. All other stabilizer generators can be obtained 
by translating this stabilizer along $x$ and $y$.
Similar to the double-chain bicycle code, the double-layer
BB codes contain two layers with an $X$ stabilizer generator 
(and similarly for $Z$ stabilizer) written as 
$S_X=X_0 X_1 X_2 X_3$ with $X_0=\prod_{j_{x1},j_{y1}} X_{0,(j_{x1},j_{y1},0)}^{a_{j_{x1} j_{y1}}}$,
$X_1=\prod_{j_{x2},j_{y2}} X_{0,(j_{x2},m-j_{y2},1)}^{b_{j_{x2},j_{y2}}}$,
$X_2=\prod_{j_{x3},j_{y3}} X_{1,(l-j_{x3},j_{y3},0)}^{a_{j_{x3},j_{y3}}}$,
and $X_3=\prod_{j_{x4},j_{y4}} X_{1,(j_{x4},j_{y4},1)}^{b_{j_{x4},j_{y4}}}$,
where the subscript includes a layer index.
All other stabilizer generators are generated by translating this
stabilizer while taking PBCs into account
[see Fig.~\ref{fig1}(a) for an example].

Table~\ref{table2} presents  
weight-eight bilayer BB codes (see more in Table~\ref{tableEM2} in End Matter) with parameters $[[n,k,d]]$ 
based on weight-four BB codes with
$A=T_x^{\alpha_1} T_y^{\beta_1}+T_x^{\alpha_2} T_y^{\beta_2}$ 
and $B=T_x^{\gamma_1} T_y^{\delta_1}+T_x^{\gamma_2} T_y^{\delta_2}$ 
(see End Matter for more codes). We observe that when both $l$ and $m$ are
even, the bilayer code is even, and when both are odd, it is 
highly probable that the code will be odd. Additionally, in cases
where one is odd and the other is even, 
an even code typically achieves the maximum 
$ kd^2/n $, although odd codes may also be found.
Among these codes listed in the table,
the maximum value of $ kd^2/n $ is $ 10.3 $ for the even-weight 
codes and $ 7.7 $ for the odd-weight codes.

\begin{table}
\begin{ruledtabular}
\centering
\caption{Weight-eight bilayer BB codes $[[n,k,d]]$  
constructed from weight-four BB codes with 
$A=x^{\alpha_1} y^{\beta_1}+x^{\alpha_2} y^{\beta_2}$ 
and $B=x^{\gamma_1} y^{\delta_1}+x^{\gamma_2} y^{\delta_2}$,
where $x=T_x$ and $y=T_y$. 
}
\label{table2}
\begin{tabular}{c c c c c c}
$[[n,k,d]]$ & $l$ & $m$ & $A$ & $B$ & $kd^2/n$  \tabularnewline
\midrule
$[[60,12,5]]$ & 3 & 5 & $x^2y^2+x^2y$ & $x^2y^2+x^2$ & 5 \tabularnewline
$[[84,8,8]]$ & 3 & 7 & $x^2y+x$ & $x^2y^2+x$ & 6.1 \tabularnewline
$[[100,12,8]]$ & 5 & 5 & $xy+x^4y$ & $y^2+x^4y^3$ & 7.7 \tabularnewline
$[[108,16,6]]$ & 9 & 3 & $x^2y^6+x^3y^3$ & $x^2y^4+x^4y$ & 5.3 \tabularnewline
$[[140,16,8]]$ & 7 & 5 & $y^4+x^2y^2$ & $y^2+x^5y$ & 7.3 \tabularnewline
\midrule
$[[56,6,8]]$ & 7 & 2 & $y^4+xy^3$ & $x^5+x^2y^3$ & 6.9 \tabularnewline
$[[80,10,8]]$ & 5 & 4 & $y^2+y$ & $x^2y+x^4$ & 8 \tabularnewline
$[[112,8,12]]$ & 14 & 2 & $x^3y^7+x^{11}y^4$ & $y^2+x^5y^{12}$ & 10.3 \tabularnewline
$[[120,8,12]]$ & 6 & 5 & $x^4+x^5y^4$ & $x+x^5y^3$ & 9.6 \tabularnewline
$[[160,20,8]]$ & 4 & 10 & $y^3+x^2y^2$ & $x+x^3 y^3$ & 8 \tabularnewline
\end{tabular}
\end{ruledtabular}
\end{table}

\emph{Laurent polynomial formalism for the bilayer BB codes}.---We now apply the theory of Laurent
polynomials to describe the bilayer BB codes. Previously, this theory has been used
to characterize the BB codes~\cite{liang2024extracting,liang2025generalized}. 
Specifically, given two polynomial $f(x,y) = \sum_{j_x,j_y} a_{j_x j_y} x^{j_x} y^{j_y}$ 
and $g(x,y) = \sum_{j_x,j_y} b_{j_x j_y} x^{j_x} y^{j_y}$ in terms of two variables $x$ and $y$, 
we represent a Pauli string using the vector
$\begin{pmatrix}
f(x,y) & g(x,y)
\end{pmatrix}$
whose support is defined by the exponents of the monomials 
in $f(x,y)$ and $g(x,y)$, corresponding to the seed stabilizer
obtained for the code with $A=f(T_x,T_y)$ and $B=g(T_x,T_y)$ as discussed in 
previous sections. Thus, we use 
$\begin{pmatrix}
f(x,y) & g(x,y)
\end{pmatrix}^T$
to represent the $X$ seed stabilizer,
from which all other $X$ stabilizers can be derived by multiplying a polynomial from 
the Laurent polynomial ring $R_{xy}=\mathbb{Z}_2[x,y,x^{-1},y^{-1}]$ 
to
$\begin{pmatrix}
f(x,y) & g(x,y)
\end{pmatrix}^T$.
Similarly, the $Z$ seed stabilizer is represented by
$\begin{pmatrix}
\overline{g(x,y)} & \overline{f(x,y)}
\end{pmatrix}^T$,
where $\overline{(\cdots)}$ denotes the antipode map, transforming all 
variables in the polynomial to their inverses, i.e., 
$x\mapsto x^{-1}$ and $y\mapsto y^{-1}$.
Within this formalism,  
the matrix $T_x$ ($T_y$) can be viewed as a matrix representation of 
$x$ of dimension $l$ ($y$ of dimension $m$) for $R_{xy}$.
For example, when $f(x,y)=1+x^{-1}$ and $g(x,y)=1+y$, we obtain the toric code.

\begin{table}[t]
\begin{ruledtabular}
\centering
\caption{Weight-eight double-layer twisted BB codes $[[n,k,d]]$ 
constructed from weight-four twisted BB codes with 
$A=x^{\alpha_1} y^{\beta_1}+x^{\alpha_2} y^{\beta_2}$ 
and $B=x^{\gamma_1} y^{\delta_1}+x^{\gamma_2} y^{\delta_2}$
with $x=\tilde{T}_x$ and $y=T_y$,
where $\tilde{T}_x$ with boundary twist $\gamma$ is provided in End Matter. }
\label{table3}
\begin{tabular}{c c c c c c c}
$[[n,k,d]]$ & $l$ & $m$ & $\gamma$ & $A$ & $B$ & $\frac{kd^2}{n}$ \tabularnewline
\midrule
$[[100,12,8]]$ & 5 & 5 & 3 & $xy^3+x^3y^3$ & $x^2y^2+x^4y^3$ & 7.7 \tabularnewline
$[[132,8,12]]$ & 3 & 11 & 9 & $1+x^2y^2$ & $y^2+x^1y$ & 8.7 \tabularnewline
$[[140,16,8]]$ & 5 & 7 & 1 & $x^2y+y$ & $x^2+x^4y^2$ & 7.3 \tabularnewline
$[[180,20,8]]$ & 5 & 9 & 4 & $y^3+x$ & $x^4y+xy^2$ & 7.1 \tabularnewline
$[[204,8,16]]$ & 17 & 3 & 2 & $x^2y^5+x^{14}y$ & $x^{11}y^{16}+x^{13}y^{10}$ & 10 \tabularnewline
\midrule
$[[112,8,12]]$ & 2 & 14 & 6 & $y+xy$ & $1+y$ & 10.3 \tabularnewline
$[[128,16,8]]$ & 4 & 8 & 4 & $xy+xy^3$ & $x^2y^2+xy^2$ & 8 \tabularnewline
$[[144,8,12]]$ & 2 & 18 & 10 & $1+y$ & $y+xy$ & 8 \tabularnewline
$[[176,10,12]]$ & 11 & 4 & 1 & $x^5y^2+x^2y^8$ & $x^{10}y+y^2$ & 8.2 \tabularnewline
$[[208,8,16]]$ & 26 & 2 & 1 & $x^{10}y^{21}+x^7y^{21}$ & $x^{10}y^4+xy^{21}$ & 9.8 \tabularnewline
\end{tabular}
\end{ruledtabular}
\end{table}

In our case, we introduce an additional variable $z$ 
satisfying that $z^2=1$ and use the polynomial $u(x,y,z)=f(x,y) + z \overline{g(x,y)}$
to describe the matrix $U$.
The variable $z$ thus serves to characterize the double layers.
For instance, let $f(x,y)=1+x$ and $g(x,y)=1+y^2$. A seed $X$ or $Z$ stabilizer 
is then expressed as 
$\begin{pmatrix}
{u(x,y,z)} & \overline{u(x,y,z)}
\end{pmatrix}^T=
\begin{pmatrix}
1+x+z(1+y^{-2}) & 1+x^{-1}+z(1+y^2)
\end{pmatrix}^T
$ as shown in Fig.~\ref{fig1}(d).
All other $X$ or $Z$ stabilizers can be obtained by multiplying an element in the Laurent
polynomial ring $R_{xyz}=\{f(x,y)+z\overline{g(x,y)}:f(x,y),g(x,y)\in R_{xy}  \}$ to
$\begin{pmatrix}
{u(x,y,z)} & \overline{u(x,y,z)}
\end{pmatrix}^T$.

\begin{figure*}[t]
\centering 
\includegraphics[width=\linewidth]{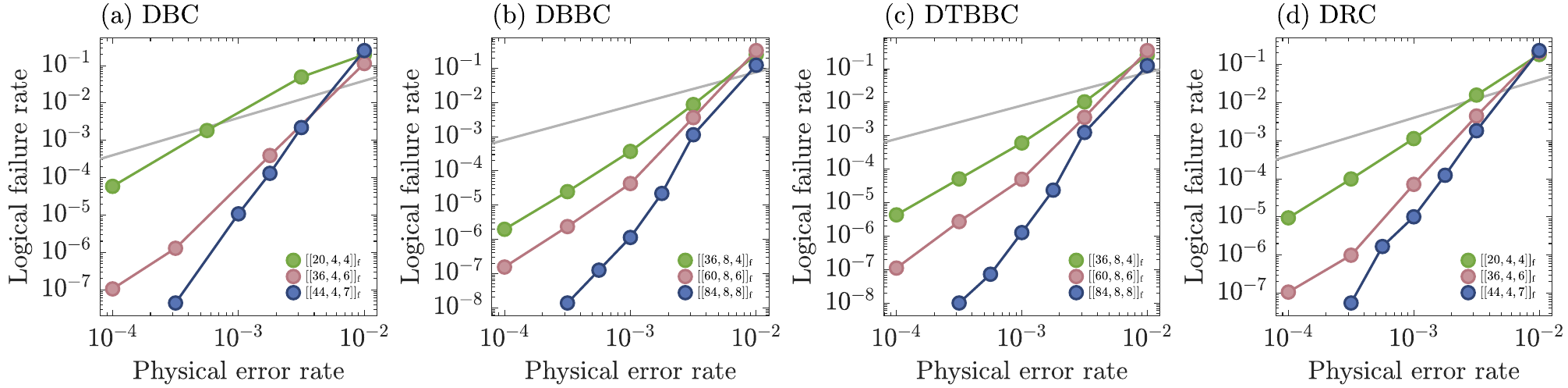}
\caption{Logical failure rate versus physical error rate under the circuit-level noise model
for (a) double-chain bicycle codes (DBC), (b) double-layer BB codes (DBBC), (c) double-layer twisted
BB codes (DTBBC), and (d) double-layer reflection codes (DRC). 
All of these codes have logical operators of odd-weight.
The grey lines represent the probability of occurrence of an error on at least 
one physical qubit for $k$ physical qubits, where $k$ is the number of logical qubits
in qubit memory.
}
\label{fig2}
\end{figure*}

We now follow Refs.~\cite{eberhardt2024logical, liang2024extracting,liang2024operator,liang2025generalized,liang2025self} to calculate the number of different types of excitations
causing syndromes.
Since our code is self-dual, it suffices to consider local Pauli $Z$ strings that
produce syndromes for $X$ stabilizers. Let 
$S_{X}=\begin{pmatrix}
u_0(x,y,z) & \overline{u_0(x,y,z)}
\end{pmatrix}^T$ with
$u_0(x,y,z) \in R_{x,y,z}$ be an $X$ seed stabilizer.
We write a local Pauli $Z$ string as 
$P=\begin{pmatrix}
	f(x,y,z) & g(x,y,z)
\end{pmatrix}^T$
, where $f(x,y,z),g(x,y,z)\in R_{xyz}$.
The syndromes generated by this Pauli string are given by 
$S_X\cdot P=\overline{u_0(x,y,z)} f(x,y,z)+u_0(x,y,z)g(x,y,z)$,
which indicates which $X$ stabilizers anticommute with this Pauli string. For example, if
$S_X\cdot P=1+x+xyz$, this indicates that the $X$ stabilizer specified by $(0,0,0)$, $(1,0,0)$, and 
$(1,1,1)$ anticommutes with the string.
The number of different types of excitations generated by $Z$ Pauli strings 
is given by 
\begin{equation} \label{Eq:numL}
k_{\text{max}}= \text{dim}({R_{xyz}}/{I_u}),
\end{equation}
where 
$I_u=\langle u,\overline{u}\rangle$ is an ideal generated by $u$ and $\overline{u}$, 
$\text{dim}({R_{xyz}}/{I_u})$ denotes the number 
of independent monomials in the quotient ring ${R_{xyz}}/{I_u}$,
which can be calculated using the algorithm in Ref.~\cite{liang2025generalized}.
Consequently, the number of logical qubits is $2k_{\text{max}}$.
For a specific lattice with width $l$ and height $m$, 
PBCs impose the relations
$x^l=1$ and $y^m=1$, modifying $I_u$ to $\langle u,\overline{u}, x^l-1, y^m-1\rangle$.

\emph{Double-layer twisted BB codes}.---We now use twisted BB codes~\cite{liang2025generalized} 
as base codes to construct double-layer twisted BB codes.
For a lattice of size $l\times m$, in contrast to the 
code with PBCs requiring that
$x^l=1$ and $y^m=1$, 
a twisted code requires that $y^m=1$ and $x^ly^{-\gamma}=1$,
where $\gamma$ is an integer.
This modification implies that, for a stabilizer, after translating it by $l$ unit cells 
along $x$, a further translation of 
$\gamma$ unit cells along $y$ is required to return 
to the original stabilizer. 
See End Matter for the matrix form of $x$ and $y$.
The number of logical qubits in the double-layer twisted BB code 
can also be computed using the Laurent polynomial formalism 
described above in Eq.~(\ref{Eq:numL}).
The only modification lies in the set of algebraic relations 
imposed by the twisted boundary conditions,
that is, $x^ly^{-\gamma}=1$, modifying $I_u$ to $\langle u,\overline{u}, x^ly^{-\gamma}-1, y^m-1\rangle$.
We provide ten representative double-layer 
twisted BB codes in Table~\ref{table3}
and more in Table~\ref{tableEM3} in End Matter.

\begin{table}
\begin{ruledtabular}
\centering
\caption{Weight-eight double-layer reflection codes $[[n,k,d]]$ 
constructed from weight-four
reflection codes. Here, $x$, $y$, $p$, and $q$ represent 
$T_x$, $T_y$, $M_x$, and $M_y$, respectively.
}
\label{table4} 
\begin{tabular}{c c c c c c}
$[[n,k,d]]$ & $l$ & $m$ & $A$ & $B$ & $\frac{kd^2}{n}$ \tabularnewline
\midrule
$[[68,4,10]]$ & 17 & 1 & $x^8 y^5q +x^9 y^{12}q$ & $x^{11}y^2+x^{15}y^{16}q$ & 5.9 \tabularnewline
$[[100,12,8]]$ & 5 & 5 & $x^2py+x^2py^4$ & $x y^2 +x^3 y$ & 7.7 \tabularnewline
$[[120,8,10]]$ & 10 & 3 & $x^2p y^4+x y^9$ & $x^7 y^4 q+x^3 y^7q$ & 6.7 \tabularnewline
$[[180,20,8]]$ & 15 & 3 & $x^7 y^8+x^2 y^{12}q$ & $x^{14} y^{12}q+x^{10} y^{10}$ & 7.1 \tabularnewline
$[[252,16,16]]$ & 7 & 9 & $x^6 y^3 q +x^5 y$ & $x y^3+x^6 y^6 $ & 16.3 \tabularnewline
\midrule
$[[64,16,8]]$ & 4 & 4 & $x y q+x p y^3q$ & $x y q +x^3 p y$ & 16 \tabularnewline
$[[96,20,8]]$ & 4 & 6 & $x^3p y^3+x y$ & $x^2 y^2+x^2 y^3$ & 13.3 \tabularnewline
$[[120,14,10]]$ & 15 & 2 & $x^{11} y^{14}q+x^7 y^{14}$ & $x^{10} y^6 +x^{11} y^{12}q$ & 11.7 \tabularnewline
$[[128,32,8]]$ & 8 & 4 & $x^2p y^6+x^6$ & $x^2 y^5 +x^6 y q$ & 16 \tabularnewline
$[[144,16,12]]$ & 18 & 2 & $x^{11}y^6+x^{15}y^{13}q$ & $x^3 q+x^{13}y^{10}$ & 16 \tabularnewline
\end{tabular}
\end{ruledtabular}
\end{table}

\emph{Double-layer reflection code}.---In the previous sections, we consider the $A$ and 
$B$ matrices, which are polynomials of either a cyclic shift 
matrix or commuting matrices $T_x$ and $T_y$.
Geometrically, all these matrices represent translation operations. We now introduce 
reflection operations $M_x$ and $M_y$, which yield the mirror images of original 
sites with respect to the $x$-normal and $y$-normal planes, respectively. 
Their matrix representations are 
$M_x = M_l\otimes I_m$ and  $M_y = I_l \otimes M_m$,
where $M_l$ is an $l \times l$ matrix whose entries are given by 
$[M_l]_{ij}=\delta_{i+j,l+1}$ for $1\leq i,j \leq l$. 
Clearly, $M_x$ and $M_y$ do not
commute with $T_x$ and $T_y$. 
The check matrix $A$ (and similarly $B$) is then modified to
$ 
A = \sum_{i,j,k,t}c_{i,j,k,t}T_x^{i} M_x^{k} T_y^{j} M_y^{t},
$
where $c_{ijkt}\in \mathbb{Z}_2$.
The constructed matrices $A$ and $B$ do not necessarily satisfy 
the conditions for base codes as described in the previous section.
Therefore, we perform a numerical search to identify check 
matrices that satisfy the base code conditions,
which are used to construct stacked codes. 
Ten codes are summarized in Table~\ref{table4} (see more in Table~\ref{tableEM4} in End Matter). Two 
of these codes can achieve a value of $kd^2/n$ as large as $16$ for 
similar-sized codes.

\emph{Numerical simulations}.---We now perform a numerical examination of
the noise resilience of the stacked qLDPC codes used as qubit memory 
under the circuit-level noise model. 
We first initialize all physical qubits in $|0\rangle$ and
then consider the depolarizing noise with the physical error rate $p$. 
Subsequently, we perfrom $N_c=d$ rounds of syndrome extraction.
The noise is modeled as the depolarizing noise channel occurring 
after each gate operation 
for both data qubits and ancilla qubits~\cite{gidney2021stim}.
Additionally, we account for classical measurement readout 
errors in the ancilla qubits.
At the end of the circuit, we perform a round of transversal measurement by 
measuring the $Z$ operators of all physical data qubits.
This transversal measurement yields the values of all $Z$ stabilizers 
and the $Z$ logical operators.
We forward all measurement results to a Tesseract decoder~\cite{beni2025tesseract} to infer the values of 
all logical $Z$ operators so as to determine whether a logical measurement value flip occurs.
For $N_\text{sample}$ samples, if $N_\text{error}$ of them 
result in incorrect predictions from the decoder, then
the logical error rate is $P_L(N_c) = {N_\text{error}}/{N_\text{sample}}$.
The logical failure rate (LFR) is defined as 
$\text{LFR}(p) = 1 - (1 - P_L)^{{1}/{N_c}}\approx P_L/N_c$~\cite{xu2024constant, bravyi2024high}, 
which roughly indicates the logical error rate
per syndrome cycle.
The pseudo-threshold is defined as the error rate $p_{0}$ 
where $\text{LFR}(p_0)=1 - (1 - p_0)^k$, meaning that the logical failure rate is equal to
the probability that at least one error occurs for $k$ physical qubits.

Figure~\ref{fig2} illustrates the logical failure rate as a function of the physical 
error rate for several stacked codes, including double-chain bicycle codes, double-layer
BB codes, double-layer twisted BB codes, and double-layer reflection
codes, all of which have logical operators of odd-weight (see End Matter
for simulation results of stacked codes featuring even-weight logical
operators). We see that the logical failure rate can be substantially suppressed,
with the pseudo-threshold exceeding $0.7\%$.

In summary, we have introduced a general method for constructing self-dual 
qLDPC codes from non-self-dual qLDPC codes. The resulting codes can 
support logical operators of odd-weight while maintaining favorable 
parameters. Numerical simulations of several such codes as 
quantum memories further demonstrate that they can achieve a high 
pseudo-threshold.

\begin{acknowledgments}
We thank F. Wei, Y.-A. Chen, Y.-F. Mao, Y. Wu, and Z. Zhou for helpful discussions. 
This work is supported by the Innovation Program for Quantum Science and Technology (Grant No. 2021ZD0301604)
and the National Natural Science Foundation of China (Grant No. 12474265 and No. 11974201).
We also acknowledge the support by center of high performance computing, Tsinghua University.
\end{acknowledgments}   

\bibliography{reference}
% \nocite{*}

\clearpage

\onecolumngrid

\begin{center}
{\textbf{End Matter}}
\end{center}

\begin{figure*}[htbp]
	\centering 
	\includegraphics[width=\linewidth]{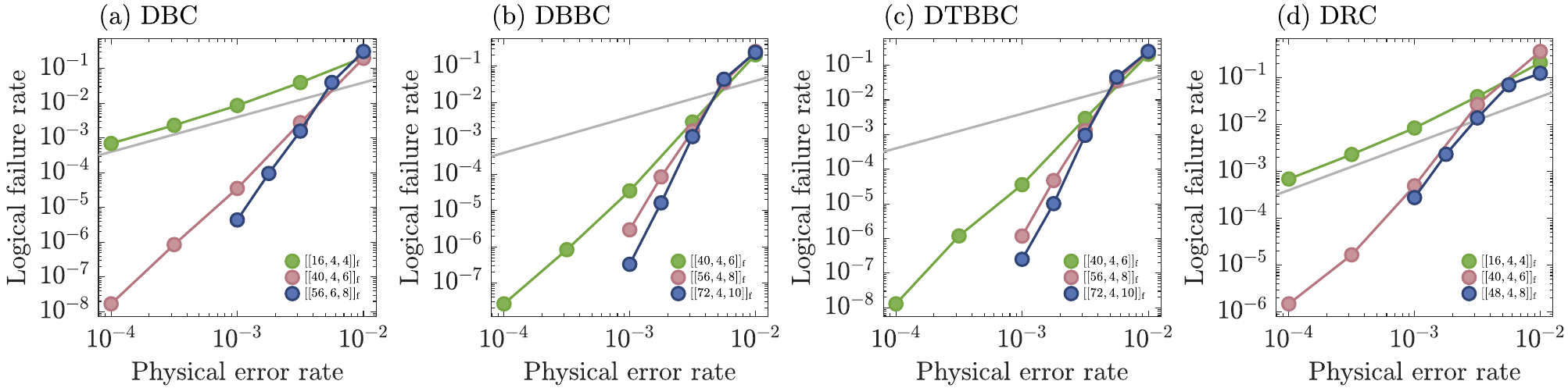}
	\caption{Logical failure rate with respect to physical error rate under the circuit-level noise model
		for (a) double-chain bicycle codes, (b) double-layer BB codes, (c) double-layer twisted
		BB codes, and (d) double-layer reflection codes. 
		All of these codes have logical operators of even-weight.
		The grey lines have the same meaning as those in Fig.~\ref{fig2}.
		The error bars hidden behind filled circles represent the standard deviation of the logical failure rate,
		$\sigma_{\text{LFR}} = (1/N_c) (1-P_L)^{\frac{1}{N_c}-1} \sigma_{P_L}$,
		where $\sigma_{P_L} = \sqrt{\frac{P_L(1-P_L)}{N_\text{sample}}}$ is 
		the standard deviation of the logical error rate $P_L$~\cite{xu2024constant, bravyi2024high}.
	}
	\label{figEM1}
\end{figure*}

\emph{More details on the double-layer twisted BB codes}.---For these codes, $y$ corresponds
to $T_y$ and $x$ corresponds to 
\begin{equation}
\tilde{T}_x=\begin{pmatrix}
0& I_m& 0  & 0& \cdots& 0\\
0&   0& I_m& 0& \cdots& 0\\
\vdots&\vdots& \vdots& \vdots& \cdots& \vdots\\
0&   0&   0& 0& \cdots& I_m\\
S_m^\gamma&   0&   0& 0& \cdots& 0\\
\end{pmatrix},
\end{equation}
which is slightly different from $T_x$.

\emph{Simulation results of even stacked qLDPC codes}.---Figure~\ref{figEM1}
presents the logical failure rate versus the physical failure rate for 
the four types of stacked codes with even-weight logical operators.
We see that the pseudo-threshold can reach up to $0.5\%$.

\begin{table*}[htbp]
\begin{ruledtabular}
\centering
\caption{More weight-eight double-chain bicycle codes, continuation of Table~\ref{table1}.
}
\label{tableEM1} 
\begin{tabular}{c c c c c c | c c c c c c}
$[[n,k,d]]$ & $l$ & $A$ & $B$ & $\frac{kd^2}{n}$ & Type & $[[n,k,d]]$ & $l$ & $A$ & $B$ & $\frac{kd^2}{n}$ & Type \tabularnewline
\midrule
$[[116,4,14]]$ & 29 & $1+x^3$ & $x^{20}+x^{25}$  & 6.8 & odd & 
$[[240,28,\leq6]]$ & 60 & $x^{41}+x^{59}$ & $x^4+x^{26}$  & 4.2 & even \tabularnewline
$[[132,8,12]]$ & 33 & $x^{10}+x^{11}$ & $x^{11}+x^{31}$  & 8.7 & odd &
$[[248,4,\leq20]]$ & 62 & $x^7+x^{26}$ & $x^{32}+x^{34}$  & 6.5 & even \tabularnewline
$[[148,4,16]]$ & 37 & $x^{21}+x^{24}$ & $x^{17}+x^{22}$  & 6.9 & odd &
$[[264,8,\leq16]]$ & 66 & $x^4+x^{18}$ & $x^{42}+x^{53}$  & 7.8 & even \tabularnewline
$[[176,16,8]]$ & 44 & $x^{11}+x^{31}$ & $1+x^8$  & 5.8 & odd &
$[[280,4,\leq22]]$ & 70 & $x^{25}+x^{51}$ & $x^{21}+x^{64}$  & 6.9 & even \tabularnewline
$[[204,4,\leq18]]$ & 51 & $1+x^{14}$ & $x^{32}+x^{40}$  & 6.4 & odd &
$[[296,6,\leq18]]$ & 74 & $x^6+x^{65}$ & $x^8+x^{25}$  & 6.6 & even \tabularnewline
$[[276,12,\leq12]]$ & 69 & $x^{27}+x^{33}$ & $x^{15}+x^{54}$  & 6.3 & odd &
$[[312,6,\leq20]]$ & 78 & $x^{14}+x^{73}$ & $x^2+x^{71}$  & 7.7 & even \tabularnewline
$[[380,12,\leq16]]$ & 95 & $x^2+x^{25}$ & $x^{58}+x^{75}$  & 8.1 & odd &
$[[360,4,\leq24]]$ & 90 & $x^{22}+x^{24}$ & $x+x^{52}$  & 6.4 & even \tabularnewline
\end{tabular}
\end{ruledtabular}
\end{table*}

\begin{table}[htbp]
\begin{ruledtabular}
\centering
\caption{More weight-eight double-layer BB codes, continuation of Table~\ref{table2}.
The codes on the right-hand side have even-weight logical operators.}
\label{tableEM2} 
\begin{tabular}{c c c c c c c|c c c c c c }
$[[n,k,d]]$ & $l$ & $m$ & $A$ & $B$ & $\frac{kd^2}{n}$ & Type & 
$[[n,k,d]]$ & $l$ & $m$ & $A$ & $B$ & $\frac{kd^2}{n}$  \tabularnewline
\midrule
$[[60,12,5]]$ & 3 & 5 & $x^2y^2+x^2y$ & $x^2y^2+x^{2}$  & 5 & odd &
$[[104,6,12]]$ & 13 & 2 & $x^2y^8+x^9y$ & $x^7y+x^6y^8$  & 8.3  \tabularnewline
$[[156,12,10]]$ & 13 & 3 & $y^6+x^{10}y^{12}$ & $x^{10}y^{11}+x^8y^8$  & 7.7 & odd &
$[[128,16,8]]$ & 16 & 2 & $x^7y^9+x^8y^9$ & $x^5y^9+x^2y^{11}$  & 8  \tabularnewline
$[[204,8,\leq16]]$ & 17 & 3 & $x+x^6y^{10}$ & $x^2y^{12}+x^6y^{11}$  & 10 & odd &
$[[168,6,16]]$ & 6 & 7 & $y^4+x^3y$ & $y^5+x^{5}$  & 9.1  \tabularnewline
$[[228,4,\leq20]]$ & 19 & 3 & $x^{17}y^{13}+x^{11}y^{13}$ & $x^{13}+x^{18}y$  & 7 & odd &
$[[192,24,8]]$ & 12 & 4 & $x^2y^3+x^7y^9$ & $x^2y^{10}+x^{11}y^2$  & 8  \tabularnewline
$[[260,20,\leq10]]$ & 5 & 13 & $x^4y^2+x^{41}$ & $x^{2}+x^2y^3$  & 7.7 & odd &
$[[208,12,\leq12]]$ & 26 & 2 & $x^{21}y^9+x^{24}y^{17}$ & $x^{20}y^4+xy^4$  & 8.3  \tabularnewline
$[[276,4,\leq22]]$ & 23 & 3 & $x^{20}y^{19}+x^{16}y^{12}$ & $x^{17}y^{18}+x^8y^9$  & 7 & odd &
$[[216,16,\leq10]]$ & 6 & 9 & $x^{2}+x^2y^5$ & $y^3+x^3y^2$  & 7.4  \tabularnewline
$[[280,32,\leq8]]$ & 7 & 10 & $xy^5+x^4y^3$ & $x^3y^3+x^6y^3$  & 7.3 & odd &
$[[224,16,\leq12]]$ & 4 & 14 & $x^3y^3+xy$ & $x^3y+x^{11}$  & 10.3 \tabularnewline
$[[364,28,\leq10]]$ & 7 & 13 & $y^5+y^4$ & $xy+xy^6$  & 7.7 & odd &
$[[248,6,\leq20]]$ & 31 & 2 & $x^{29}y^{25}+x^3y^{22}$ & $x^{29}y^{28}+x^{22}y^{11}$  & 9.7  \tabularnewline
$[[24,8,4]]$ & 3 & 2 & $xy^2+x^{2}$ & $x^2y+xy^2$  & 5.3 & even &
$[[272,8,\leq20]]$ & 17 & 4 & $x^9y^{11}+x^{10}y^8$ & $x^{31}+xy^{11}$  & 11.8  \tabularnewline
$[[32,12,4]]$ & 2 & 4 & $xy+x$ & $x+y$  & 6 & even &
$[[288,12,\leq16]]$ & 9 & 8 & $x^7y^8+x^4y^3$ & $x^8y^2+x^7y^7$  & 10.7  \tabularnewline
$[[64,24,4]]$ & 4 & 4 & $x^3y^3+x$ & $xy^2+xy^3$  & 6 & even &
$[[320,16,\leq16]]$ & 10 & 8 & $x^7y^8+x^9y^9$ & $x^5y^6+y$  & 12.8  \tabularnewline
$[[88,4,12]]$ & 11 & 2 & $y^7+xy$ & $x^9y^4+x^7y^7$  & 6.5 & even &
$[[384,48,\leq8]]$ & 12 & 8 & $xy^4+xy^{10}$ & $x^5y^{11}+x^8y^8$  & 8 \tabularnewline
\end{tabular}
\end{ruledtabular}
\end{table}

\begin{table}[htbp]
\begin{ruledtabular}
\centering
\caption{More weight-eight double-layer twisted BB codes, continuation of Table~\ref{table3}.
The codes on the right-hand side have even-weight logical operators.}
\label{tableEM3} 
\begin{tabular}{c c c c c c c c|c c c c c c c }
$[[n,k,d]]$ & $l$ & $m$ & $\gamma$ & $A$ & $B$ & $\frac{kd^2}{n}$ & Type & 
$[[n,k,d]]$ & $l$ & $m$ & $\gamma$ & $A$ & $B$ & $\frac{kd^2}{n}$ \tabularnewline
\midrule
$[[60,12,5]]$ & 3 & 5 & 4 & $x^{2}+x^2y$ & $1+y^2$  & 5 & odd &
$[[56,6,8]]$ & 7 & 2 & 1 & $xy^5+x^6y^3$ & $x^4y^4+x^5y^4$  & 6.9  \tabularnewline
$[[84,8,8]]$ & 3 & 7 & 4 & $1+xy^2$ & $y^2+x^2y^2$  & 6.1 & odd &
$[[64,8,8]]$ & 2 & 8 & 5 & $y+x$ & $xy+1$  & 8  \tabularnewline
$[[100,12,8]]$ & 5 & 5 & 3 & $xy^3+x^3y^3$ & $x^2y^2+x^4y^3$  & 7.7 & odd &
$[[72,4,10]]$ & 3 & 6 & 2 & $xy+x^{2}$ & $y^2+x^2y^2$  & 5.6  \tabularnewline
$[[220,12,\leq12]]$ & 11 & 5 & 2 & $x^9y^9+x^2y^2$ & $x^{7}+x^9y^5$  & 7.9 & odd &
$[[80,10,8]]$ & 10 & 2 & 1 & $x^9y^3+y^8$ & $x^{5}+x^2y^9$  & 8  \tabularnewline
$[[228,8,\leq16]]$ & 3 & 19 & 1 & $y^2+x$ & $1+x^2y^2$  & 9.0 & odd &
$[[104,6,12]]$ & 2 & 13 & 6 & $xy+1$ & $x+y$  & 8.3 \tabularnewline
$[[252,4,\leq20]]$ & 7 & 9 & 4 & $x^2y^4+x^{2}$ & $x^3y+x^6y^4$  & 6.3 & odd &
$[[120,8,12]]$ & 3 & 10 & 7 & $x^{2}+x$ & $x^2y+xy^2$  & 9.6   \tabularnewline
$[[260,4,\leq21]]$ & 5 & 13 & 10 & $x^4y+x^2y^4$ & $x^{4}+1$  & 6.8 & odd &
$[[136,6,12]]$ & 2 & 17 & 12 & $x+y$ & $x+1$  & 6.4  \tabularnewline
$[[324,8,\leq20]]$ & 3 & 27 & 9 & $x^2y+y^2$ & $xy^2+x^2y^2$  & 9.9 & odd &
$[[160,20,8]]$ & 4 & 10 & 1 & $x^2y+y^2$ & $x^2y^2+1$  & 8  \tabularnewline
$[[340,4,\leq22]]$ & 5 & 17 & 14 & $y^4+y^3$ & $x^{4}+x^3y$  & 5.7 & odd &
$[[216,12,\leq12]]$ & 6 & 9 & 3 & $x^5y^3+x^5y^4$ & $y^3+x^5y^2$  & 8  \tabularnewline
$[[372,8,\leq18]]$ & 3 & 31 & 16 & $xy^2+y^2$ & $x^2y^2+y$  & 7.0 & odd &
$[[224,16,\leq12]]$ & 4 & 14 & 2 & $x^3y^2+xy$ & $x^3y^3+x^{3}$  & 10.3  \tabularnewline
$[[24,8,4]]$ & 3 & 2 & 1 & $y^2+x^2y$ & $y+x$  & 5.3 & even &
$[[240,16,\leq12]]$ & 6 & 10 & 2 & $y^3+x^5y$ & $x^4y^2+y^5$  & 9.6  \tabularnewline
$[[32,12,4]]$ & 2 & 4 & 1 & $xy+1$ & $x+1$  & 6 & even &
$[[288,12,\leq16]]$ & 8 & 9 & 1 & $x^5y^4+y^3$ & $x^5y^2+x^2y^5$  & 10.7  \tabularnewline
$[[48,16,4]]$ & 2 & 6 & 2 & $y+xy$ & $xy+x$  & 5.3 & even &
$[[384,8,\leq20]]$ & 3 & 32 & 6 & $y+xy^2$ & $x^2y+xy^2$  & 8.3 \tabularnewline
\end{tabular}
\end{ruledtabular}
\end{table}

\begin{table}[h]
\begin{ruledtabular}
\centering
\caption{More weight-eight double-layer reflection codes, continuation of Table~\ref{table4}.
The codes on the left-hand and right-hand sides have odd-weight and even-weight logical 
operators, respectively.}
\label{tableEM4} 
\begin{tabular}{c c c c c c |c c c c c c }
$[[n,k,d]]$ & $l$ & $m$ & $A$ & $B$ & $\frac{kd^2}{n}$  &
$[[n,k,d]]$ & $l$ & $m$ & $A$ & $B$ & $\frac{kd^2}{n}$ \tabularnewline
\midrule
$[[28,4,5]]$ & 7 & 1 & $x^2y^2+x^5y^6q$ & $x^3y^3q+xy^6q$  & 3.6  &
$[[24,8,4]]$ & 3 & 2 & $y^2q+xyq$ & $x^2y+q$  & 5.3  \tabularnewline
$[[36,12,4]]$ & 3 & 3 & $pyq+x$ & $pq+xy^2q$  & 5.3 &
$[[32,18,4]]$ & 4 & 2 & $x^3y^3+xpy^3q$ & $x^3py^3+x^3$  & 9 \tabularnewline
$[[44,4,7]]$ & 11 & 1 & $x^9y^8+x^2y^3q$ & $xy^2+x^7y^2$  & 4.5 &
$[[96,20,8]]$ & 4 & 6 & $x^3py^3+xy$ & $x^2y^2+x^2y^3$  & 13.3 \tabularnewline
$[[60,12,5]]$ & 15 & 1 & $y^2+x^6y^{10}$ & $x^{10}y^9q+x^7y^7$  & 5 &
$[[112,8,14]]$ & 14 & 2 & $x^5y^3+x^6q$ & $x^{10}y^{12}q+x^5y^5q$  & 14 \tabularnewline
$[[132,4,18]]$ & 11 & 3 & $x^7y^2+x^6$ & $x^{10}y^{10}q+x^7yq$  & 9.8 &
$[[120,14,10]]$ & 15 & 2 & $x^{11}y^{14}q+x^7y^{14}$ & $x^{10}y^6+x^{11}y^{12}q$  & 11.7  \tabularnewline
$[[260,4,\leq26]]$ & 13 & 5 & $x^{12}y^7+x^7y^3$ & $x^{10}y^2q+x^{11}y^9q$  & 10.4 &
$[[160,8,16]]$ & 10 & 4 & $x^2y^8+x^6y$ & $x^7y^3q+y^5q$  & 12.8 \tabularnewline
$[[300,4,\leq28]]$ & 25 & 3 & $x^3y^5+x^{24}y^4$ & $x^{14}y^6+x^{19}y^2q$  & 10.5 &
$[[200,4,\leq20]]$ & 25 & 2 & $x^{15}y^2q+x^2y^{15}q$ & $x^4y^{22}q+x^{13}y^4q$  & 8  \tabularnewline
$[[324,4,\leq28]]$ & 27 & 3 & $x^2y^4+x^6y^{17}$ & $x^{22}y^{17}q+x^{16}y^{14}q$  & 9.7 &
$[[256,8,\leq20]]$ & 32 & 2 & $x^{28}y^6+x^9y^{22}q$ & $x^{13}y^{26}q+x^{15}y^6q$  & 12.5 \tabularnewline
$[[348,4,\leq35]]$ & 29 & 3 & $x^{21}y^5+x^{24}y^{10}q$ & $x^{27}y^{19}q+x^{22}y^{20}$  & 14.1 &
$[[312,6,\leq30]]$ & 13 & 6 & $x^{11}y^7q+y^5$ & $xy^5+x^5y^7q$  & 17.3 \tabularnewline
$[[380,4,\leq58]]$ & 5 & 19 & $x^4y^3+x^3py^4$ & $xy+x^4y^3$  & 35.4 &
$[[384,24,\leq16]]$ & 24 & 4 & $x^{16}y^8+x^5y^2$ & $x^{13}y^{11}q+x^{10}y^{11}$  & 16 \tabularnewline
\end{tabular}
\end{ruledtabular}
\end{table}

\end{document}